\documentclass{article}

\usepackage{graphicx}
\usepackage{pifont}
\usepackage{latexsym}
\usepackage{amsfonts}
\usepackage{amsmath}
\usepackage{amssymb}
\usepackage{commath}
\usepackage{float}
\usepackage{tikz}
\usepackage{enumitem}
\usepackage{qcircuit}
\usepackage{caption}
\usepackage{subcaption}
\usepackage{csquotes}

\usepackage[affil-it]{authblk}
\usepackage[english]{babel}
\usepackage{blindtext}

\usepackage{cite}

\usepackage{tikz}
\usetikzlibrary{arrows,pgfplots.groupplots,external}
\usepackage{pgfplots}
\pgfplotsset{compat=1.10}
\usepackage[detect-family]{siunitx}
\usepackage[eulergreek]{sansmath}
\usepackage{relsize}

\newcommand{\targalone}{*+<.02em,.02em>{\xy ="i","i"-<.39em,0em>;"i"+<.39em,0em> **\dir{-}, "i"-<0em,.39em>;"i"+<0em,.39em> **\dir{-},"i"*\xycircle<.4em>{} \endxy}}

\newcommand{\gateShorthand}[2]{{\Qcircuit @R=1em @C=0.75em @!R=0.5em {
   \push{\phantom{#2}[#1]#2}\qwx[1]{1} \\
   \targalone \\
}}}
\newcommand{\cleanGate}[1]{\gateShorthand{#1}{\phantom{*}}}
\newcommand{\dirtyGate}[1]{\gateShorthand{#1}{*}}

\begin{document}
\sloppy

\title{Decomposing Quantum Generalized Toffoli with an Arbitrary Number of Ancilla}

\author{Jonathan M. Baker\footnote{jmbaker@uchicago.edu}, Casey Duckering\footnote{cduck@uchicago.edu}, Alexander Hoover\footnote{alexhoover@uchicago.edu}, Frederic T. Chong\footnote{chong@cs.uchicago.edu} \\ University of Chicago Department of Computer Science}

\date{March 27, 2019}

\maketitle

\begin{abstract}
\noindent We present a general decomposition of the Generalized Toffoli, and for completeness, the multi-target gate using an arbitrary number of clean or dirty ancilla. While prior work has shown how to decompose the Generalized Toffoli using 0, 1, or $O(n)$ many clean ancilla and 0, 1, and $n-2$ dirty ancilla, we provide a generalized algorithm to bridge the gap, i.e. this work gives an algorithm to generate a decomposition for any number of clean or dirty ancilla. While it is hard to guarantee optimality, our decompositions guarantee a decrease in circuit depth as the number of ancilla increases.
\end{abstract}

\section{Introduction}

 

Classical operations such as \texttt{AND} and \texttt{OR} are inherently irreversible. However, they can be implemented reversibly via the Toffoli gate, a three-input three-output gate which performs a \texttt{NOT} (or X) gate on the target conditioned on the other two inputs, the controls, being on. The Toffoli, and its generalized version the Generalized Toffoli gate (sometimes abbreviated as $C^nX$ or $C^nU$ when another gate other than \texttt{NOT} is to be applied), is an important operation in quantum computation, an emerging technology promising to provide as much as exponential speedup to classically hard problems.

In quantum computation, however, we cannot natively perform these large gates on hardware. Instead, we must first decompose these gates into a sequence of one and two qubit (quantum bit) gates. There are two primary goals in these decompositions: low depth and low gate count, where depth is roughly equivalent to run time and can be determined by the length of the critical path through the circuit.

One way to improve both of these metrics is by the inclusion of ancilla, temporary scratch bits used with the promise to return to their original state at the end of the computation. There are two main classes of ancilla used: clean and dirty. A clean ancilla is one which is used and the initial state is known, typically 0. A dirty ancilla arrives in an unknown state and usually requires additional gates to take advantage of the additional space. In both cases they must be returned to the original state via a process of \textit{uncomputation}, essentially undoing a sequence of operations performed on it. 

Both of these classes have been explored previously to improve decompositions of the Generalized Toffoli. Notably, Gidney has provided constructions for the in-place (0 ancilla) Toffoli which is made possible by the existence of roots of all quantum gates. Gidney has also provided a 1 clean ancilla decomposition and an $n-2$ dirty ancilla decomposition \cite{GidneyBlogPost}. He et al. provide a decomposition using a linear number of clean ancilla \cite{He}. Finally, Barenco gave a quadratic depth, 0 ancilla construction; however, we do not use this construction \cite{Barenco}.

In this paper we present how to construct a decomposition of the Generalized Toffoli given any fixed number, $0 \le k \le n-2$, of clean or dirty ancilla. Specifically, we bridge the gap between these prior works and provide low depth decompositions for the whole range of possible ancilla. Finally, we consider the multi-target gate for completeness, i.e. a gate with a single control and many targets.

\section{The Generalized Toffoli, $C^nU$ Gate}
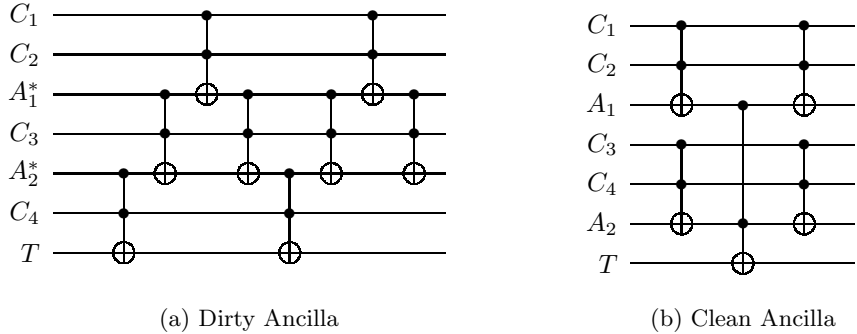
\begin{figure}%
    \centering
    \begin{subfigure}[b]{0.4\textwidth}
        \[\Qcircuit @R=1em @C=0.75em @!R=0.5em {
            \lstick{C_1} & \qw & \qw & \qw      & \qw     & \ctrl{2}& \qw     & \qw      & \qw     & \ctrl{2}& \qw &\qw \\
            \lstick{C_2}&\qw & \qw & \qw      & \qw     & \ctrl{1}& \qw     & \qw      & \qw     & \ctrl{1}& \qw &\qw \\
            \lstick{A_1^*} & \qw & \qw & \qw      & \ctrl{2}& \targ   & \ctrl{2}& \qw      & \ctrl{2}& \targ   & \ctrl{2}&\qw \\
            \lstick{C_3}& \qw & \qw & \qw      & \ctrl{1}& \qw     & \ctrl{1}& \qw      & \ctrl{1}& \qw     & \ctrl{1}&\qw \\
            \lstick{A_2^*}& \qw & \qw & \ctrl{2} & \targ   & \qw     & \targ   & \ctrl{2} & \targ   & \qw     & \targ&\qw \\
            \lstick{C_4}& \qw & \qw & \ctrl{1} & \qw     & \qw     & \qw     & \ctrl{1} & \qw     & \qw     & \qw&\qw \\
            \lstick{T}&   \qw & \qw & \targ    & \qw     & \qw     & \qw     & \targ    & \qw     & \qw     & \qw&\qw \\
        }\]
        \caption{Dirty Ancilla}
    \end{subfigure}
    \quad
    \begin{subfigure}[b]{0.4\textwidth}
            \[\Qcircuit @R=1em @C=0.75em @!R=0.5em {
                    \lstick{C_1} &\qw& \ctrl{2} &\qw& \qw       &\qw& \ctrl{2} & \qw&\qw \\
                    \lstick{C_2} &\qw& \ctrl{1} &\qw& \qw       &\qw& \ctrl{1} & \qw&\qw \\
                    \lstick{A_1} &\qw& \targ    &\qw& \ctrl{4}  &\qw& \targ & \qw&\qw \\
                    \lstick{C_3} &\qw& \ctrl{2} &\qw& \qw       &\qw& \ctrl{2} & \qw&\qw \\
                    \lstick{C_4} &\qw& \ctrl{1} &\qw& \qw       &\qw& \ctrl{1} & \qw&\qw \\
                    \lstick{A_2} &\qw& \targ    &\qw& \ctrl{1}  &\qw& \targ & \qw&\qw \\
                    \lstick{T}   &\qw& \qw      &\qw& \targ     & \qw & \qw&\qw&\qw \\
                }\]
        \caption{Clean Ancilla}
    \end{subfigure}
    \caption{The two main components, from \cite{GidneyBlogPost} (a) and \cite{He} (b). In both we have 4 controls, a single target, and two ancilla. In (a) we use two dirty ancilla $A_1^*$ and $A_2^*$ while in (b) we rely on two clean ancilla $A_1$ and $A_2$. In both, we restore the ancilla to their original states.}
    \label{fig:circ_components} 
\end{figure}

In our Generalized Toffoli decompositions, we will mainly consider Toffoli depth, i.e. we first decompose to a circuit comprised first of Toffoli gates and CNOT gates. There are well known decompositions of the Toffoli which we can then use to generate the final decomposition using only single and two qubit gates \cite{ToffoliDecomp}. The basic subcircuits of our decompositions can be found in Figure \ref{fig:circ_components}. 

\begin{figure}[!ht]
    \centering
    \begin{subfigure}[b]{0.4\textwidth}
        \[\Qcircuit @R=1em @C=0.4em @!R=0.5em {
        \lstick{} & \push{[n, m]}\qwx[1]{1} & \\
        \lstick{} & \targ & \qw \\
        }\]
        \caption{$m$ clean ancilla}
    \end{subfigure}
    \quad
    \begin{subfigure}[b]{0.4\textwidth}
        \[\Qcircuit @R=1em @C=0em @!R=0.5em {
        \lstick{} & \push{{\phantom{*}}[n, m]*}\qwx[1]{1} & \\
        \lstick{} & \targ & \qw \\
        }\]
        \caption{$m$ dirty ancilla}
    \end{subfigure}
    \caption{A Generalized Toffoli with $n$ controls and $m$ either clean or dirty ancilla. The target is assumed to be on a clean ancilla, rather than on the target unless specified.}
    \label{fig:notation}
\end{figure}
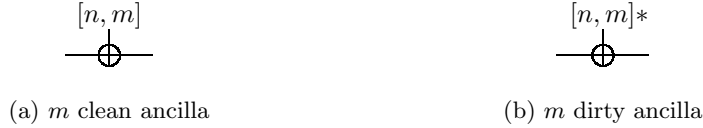

For simplicity of notation in circuits, we will introduce a concise version of the $C^nU$ gate given as in Figure \ref{fig:notation}. We never consider a construction using both dirty and clean ancilla at the same time, though the target of a dirty construction may be a clean ancilla. As we will see, our schemes opt at every step to use either clean or dirty constructions but not both. 

\subsection{Clean Ancilla}
The basis for decomposing the Generalized Toffoli gate with an arbitrary number of ancilla are Gidney's $n-2$ dirty ancilla version and He et. al.'s log-depth clean ancilla version, as noted previously. This decomposition can also be seen as a generalization of the 1 clean ancilla decomposition.

We build on these by introducing a dynamic programming approach to generating a decomposition. To optimize for depth, we bound the Toffoli depth of our decomposition with $n$ controls and $m$ clean ancilla and denote this bound $D(n,m)$, which we solve for using dynamic programming. By reconstructing the solution, we can find the actual decomposition. Note, $D(n,m)$ is not the true Toffoli depth of the circuit, but our bound on it. What we take as our base cases are the following:
\begin{align}
    D(0, m) &= D(1, m) = 0 &(m \ge 0) \\
    D(2,m) &= 1 &(m\ge 0)\\
    D(n,m) &= 2\cdot\lceil \log n\rceil - 1 &(m\ge n-2)\\
    D(n,0) &= 4n - 8 &(n\ge 3)
\end{align}
where (1) is simply a CNOT gate, (2) is a Toffoli gate, (3) comes from \cite{He}, and (4) comes from \cite{GidneyBlogPost}.

For the last of our base cases, we assume we have access to $n-2$ dirty ancilla, which we will guarantee via our dynamic programming scheme. This means we may assume our initial input gives $m\ge 1$. In the case where $m=0$, we simply return the 0 ancilla decomposition of \cite{GidneyBlogPost}. To find the bounded Toffoli depth of a decomposition with $n$ controls and $m$ clean ancilla, we need to reduce our instance size to recursively find a solution. To do this, we consider using our ancilla in two ways: as intermediate targets or as catalysts to speed up simple subcircuits.

\subsubsection*{\textbf{Our Two Schemes at a High Level}}
To accomplish this, we have two schemes. The first scheme breaks the problem into multiple log-depth clean ancilla subcircuits which store the \texttt{AND} of their set of controls on some clean ancilla. In this scheme, we give the log-depth circuits ancilla which are reclaimed later. We consider number of controls and ancilla to determine how many of these subcircuits we can run in parallel without exceeding our number of ancilla. We also consider multiple ways to divide our controls into these subcircuits to break up our circuit in the best way possible. By dividing our controls and ancilla into $p$ subcircuits, then after performing the subcircuits, we will be left with the subproblem $D(n-c+p, m-p)$, where $c$ is the number of controls that were assigned to one of the $p$ subcircuits.
In this case, we ensure there are enough ancilla to run our log-depth clean ancilla subcircuits.

In our second scheme, we make use of the $n-2$ dirty ancilla decomposition (base case 4). In theory, this scheme works as a ``backup'' to the previous scheme and should usually be taken only when there aren't enough ancilla to efficiently do log-depth subcircuits. For this method of decomposing, we break some of our controls into groups. These groups decompose to the $n-2$ dirty ancilla decomposition using the unassigned controlled ancilla as dirty ancilla and use a free clean ancilla as a target. This is similar in character to the one clean ancilla decomposition; however, we generalize this idea to allow more than one clean ancilla. Notice, after breaking into $p$ groups of dirty ancilla subcircuits, we are left with the subproblem $D(n-c+p,m-p)$, where $c$ again is the number of controls assigned to one of the $p$ subcircuits. In this case, we want to guarantee there are enough controls in the system to be used as dirty qubits when we break into smaller groups.

\subsubsection*{\textbf{Scheme 1, In Detail}}
We now describe how to break our controls into groups for the first scheme. We either want to be using every ancilla available and/or assign every control qubit to one of the log-depth constructions. So, for every $2\le k\le m+1$, we consider $k$ to be the size of the largest subcircuit allowed. Then, we assign qubits and controls to subcircuits of size $k$, until there are no longer enough ancilla or controls. We then consider allowing circuits of size $k-1$, then $k-2$, etc. We repeat this process for each $k$ in order to explore both decomposing into a large subcircuit or into many smaller subcircuits (Toffoli gates at the smallest). Notice, for each of these subcircuits of size $k$, we require $k-1$ ancilla rather than $k-2$, because we must write to some temporary target.

\begin{figure}
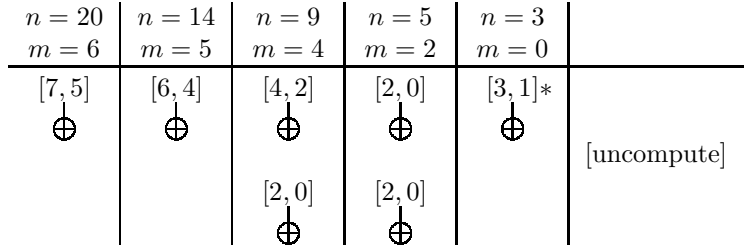

    \centering

\begin{tabular}{c|c|c|c|c|c}
    $n=20$ & $n=14$ & $n=9$ & $n=5$ & $n=3$ & \\
    $m=6$ & $m=5$ & $m=4$ & $m=2$ & $m=0$ & \\
    \hline
    \cleanGate{7,5} & \cleanGate{6,4} & \cleanGate{4,2} & \cleanGate{2,0} & \dirtyGate{3,1} & \\
    &&&&& [uncompute] \\
                    &                 & \cleanGate{2,0} & \cleanGate{2,0} &                 & \\
\end{tabular}
    \caption{Decomposing a Generalized Toffoli with $n=20$ controls and $m=6$ ancilla via our two scheme process. Each column represents a layer of the decomposition where we choose either clean or dirty constructions. Our dynamic programming algorithm decides on what to do in each layer. Once we have 0 free clean ancilla, we use scheme 2 onto the original target. Afterwards, we perform uncomputation to restore ancilla.} 
    \label{fig:clean_decomp}
\end{figure}

For each of these possible decompositions, we consider our \textit{cost} to be the depth of the largest subcircuit or $2\cdot(2\cdot\lceil \log k\rceil - 1)$ (factor of two due to uncomputation to restore the ancilla for the next layer). Therefore, we select $k$ to minimize $2\cdot(2\cdot\lceil\log k \rceil - 1) + D(n-c+p,m-p)$, where $c$ is the number of controls assigned to one of the subcircuits and $p$ is the number of subcircuits. The values of $p$ and $c$ are explicitly calculated from the given $k$ and are unique because we greedily form subcircuits. In the case where $p=m$, we check if there are enough dirty ancilla in the circuit to perform $n-2$ dirty ancilla subcircuit from our base case. If impossible, we do not generate the outcome as a possible decomposition.

\subsubsection*{\textbf{Scheme 2, In Detail}}
There are cases when the best the above scheme can do is far from optimal or will yield no solution. For example, given a single ancilla, the above scheme will always do a single Toffoli gate onto the ancilla then attempt the remainder of the circuit (often without enough dirty ancilla for our base case). In these cases, we certainly need a fallback scheme that will always work. For comparison, we also test this scheme in all cases; however, it does not improve over the first scheme except with few ancilla.

We will now describe how to split in scheme 2. Here, we want to maximize the number of $n-2$ dirty ancilla constructions in parallel, while ensuring there are an appropriate number of usable dirty ancilla. Finally, we try to use all of the ancilla in this scheme, while ensuring there are enough dirty ancilla in the next step to perform the $n-2$ dirty ancilla construction onto the final target. 

The first of these criteria give use the restricting inequality $n-m\cdot k\ge m\cdot(k-2)$,  stating, the number qubits not assigned to one of the $m$ groups of $k$ controls must be enough to serve as dirty ancilla to the groups. The second restriction gives us the inequality $m\cdot k\ge (n-m\cdot k) + m - 2$, stating the ancilla assigned to groups in this step must be enough to serve as the dirty ancilla for the dirty ancilla construction onto the target. Together this gives
\begin{align}
    \frac{n+m-2}{2m} \le k \le \frac{n + 2\cdot m}{2m}
\end{align}

Finally, we select $k$ to optimize for depth. We have two layers of $n-2$ dirty ancilla subcircuits. The depths of the layers respectively are $4k-8$ and $4(n-m\cdot k+m)-8$. However, to uncompute, we will have to pay for the first layer twice. So, we select the integer $k$ in the above interval minimizing $(2-m)k$ (smallest possible k if $m=1$ and largest possible k if $m\ge2$). The final cost of this scheme is $2\cdot(4\cdot k-8)+D(n-m\cdot k + m, 0)$.

There are some cases where no integer lies in the interval above. When this happens, we test $p=m-1,m-2,\ldots,1$ and search for an integer in
\begin{align}
    \frac{n+p-2}{2p} \le k \le \frac{n + 2\cdot p}{2p}
\end{align}
Once we find an integer $k$ in that lies in one of these intervals, we split into $p$ groups. And pay a cost of $2\cdot(4\cdot k - 8) + D(n-p\cdot k + p, m-p)$. In general, these cases happen when there are too many ancilla in the system, so they aren't usually selected over scheme 1. However, we include this completion of scheme 2 as a ``backup'', in case scheme 1 and the first step of scheme 2 do not yield decompositions.

Together, our first and second scheme give an approach with dynamic programming to decompose any Generalized Toffoli gate given an arbitrary number of clean ancilla. An example of our dynamic programming schemes can be found in Figure \ref{fig:clean_decomp}.

\subsection{Dirty Ancilla}

We also provide a decomposition in which none of the provided ancilla are clean. In this case, it must be noted that $U$ mus  be a self-inverse. With no ancilla, we give the same in-place decomposition as with 0 clean ancilla. With fewer than 3 controls, we can get no improvement with the use of dirty ancilla, and simply return a CNOT for a single control and a Toffoli for two controls. For a number of dirty ancilla $1 \le m \le n - 2$ where $n$ is the number of controls, it is illustrative to examine Gidney's decomposition when $m = n - 2$, as in Figure \ref{fig:circ_components}(a). This decomposition relies heavily on the bit toggling trick, a primary method of using dirty ancilla, which essentially means you do the circuit twice.

There are two sections in this decomposition, the up and down, each repeated twice, as labeled (up, down, up, down). We think of every set of controls as a group each with a single dirty ancilla as the target, except for the group which controls onto the target of the original gate. We can then suppose one of these ancilla has been deleted, and we then have one fewer groups. The controls of the deleted group must then be relocated to another group. Doing so will necessarily require the use of a Generalized Toffoli for that group. For every Generalized Toffoli gate we use with $k < n$ controls, we ensure there are enough dirty qubits available to perform the decomposition with $k-2$ dirty ancilla, i.e. every group must have size $k$ such that $2k - 2 < n$. 

The Generalized Toffoli with $n$ controls and $n-2$ dirty ancilla has Toffoli depth $4n - 8$, for $n \ge 3$. The Toffoli depth then of a $C^aX$ followed by a $C^bX$, with $a > b > 2$, with the requisite number of dirty ancilla has depth equivalent to a $C^{a-i}X$ followed by $C^{b+i}X$, again with the requisite number of dirty ancilla, for all $i < a - 2$. However, the depth of $C^2X$ followed by a $C^aX$ for $a > 2$ does not have depth equivalent to $C^{2+i}X$ followed by $C^{a-i}X$ for $i < a-2$, in fact this has greater depth, implying it is more efficient to use as many Toffoli gates as possible as we have fewer and fewer dirty ancilla.

Together, the above suggests the following facts about how this decomposition should operate:

\begin{enumerate}
    \item The number of controls in a group should not be so big that it prohibits the use of the $n-2$ dirty ancilla construction internally.
    \item We should preserve as many Toffoli gates as possible.
\end{enumerate}

Suppose we have a Generalized Toffoli with $n$ controls. If we have 0, ancilla we return the 0 ancilla construction as we did with 0 clean ancilla. If we have more than $n-2$ ancilla, we restrict ourselves to using $n-2$ ancilla. Suppose then we have $1 \le m \le n-2$ dirty ancilla. Let $C_1, ..., C_n$ be the $n$ controls, let $T$ be the target, let $A_1^*, ..., A_m^*$ be the dirty ancilla provided. We begin by constructing $m+1$ lists $G_1, ..., G_{m+1}$. We then append to $G_i$ $A_{i-1}^*$ for $i > 1$. Then, we begin by filling each group with exactly $2$ controls, i.e. append to $G_{i}$ $C_{2i}$ and $C_{2i+1}$ for each $i$. Then, if there are any remaining controls, begin with $G_1$ and append controls as long as $\abs{G_1} - 1 = k < n/2 + 1$ (via fact 1). Because of this setup, we then append to $G_{m+1}$ in the same fashion and then go in order $G_{2}, ..., G_{m}$. The reason for this is because the top and bottom groups appear only twice in the decomposition, i.e. they only appear in the up sections. Groups in the middle appear in both the up and down sections, and therefore are repeated 4 times. Finally, append $A_i^*$ to $G_i$ for each $1 \le i \le m$ and append $T$ to $G_{m+1}$.

\begin{figure}
    \centering
    \begin{subfigure}[b]{0.4\textwidth}
        \[\Qcircuit @R=1em @C=0.6em @!R=0.5em {
            \lstick{C_1}    &\qw     &\ustick{\text{Up 1}}\qw     &\qw      &\ctrl{2}&\qw       &\ustick{\text{Down 1}}\qw     &\qw     &\qw      &\qw        &\ctrl{2} &\qw      &\qw&\qw\\
            \lstick{C_2}    &\qw     &\qw     &\qw      &\ctrl{1}&\qw       &\qw     &\qw     &\qw      &\qw        &\ctrl{1} &\qw      &\qw&\qw\\
            \lstick{A_1^*}   &\qw     &\qw     &\ctrl{2} &\targ   &\ctrl{2}  &\qw     &\qw     &\qw      &\ctrl{2}   &\targ    &\ctrl{2} &\qw&\qw\\
            \lstick{C_3}    &\qw     &\qw     &\ctrl{1} &\qw     &\ctrl{1}  &\qw     &\qw     &\qw      &\ctrl{1}   &\qw      &\ctrl{1} &\qw&\qw\\
            \lstick{A_2^*}   &\qw     &\ctrl{2}&\targ    &\qw     &\targ     &\ctrl{2}&\qw     &\ctrl{2} &\targ      &\qw      &\targ    &\ctrl{2}&\qw\\
            \lstick{C_4}    &\qw     &\ctrl{1}&\qw      &\qw     &\qw       &\ctrl{1}&\qw     &\ctrl{1} &\qw        &\qw      &\qw      &\ctrl{1}&\qw\\
            \lstick{A_3^*}   &\ctrl{4}&\targ   &\qw      &\qw     &\qw       &\targ   &\ctrl{4}&\targ    &\qw        &\qw      &\qw      &\targ&\qw\\
            \lstick{C_5}    &\ctrl{3}&\qw     &\qw      &\qw     &\qw       &\qw     &\ctrl{3}&\qw      &\qw        &\qw      &\qw      &\qw&\qw\\
            \lstick{C_6}    &\ctrl{2}&\qw     &\qw      &\qw     &\qw       &\qw     &\ctrl{2}&\qw      &\qw        &\qw      &\qw      &\qw&\qw\\
            \lstick{C_7}    &\ctrl{1}&\qw     &\qw      &\qw     &\qw       &\qw     &\ctrl{1}&\qw      &\qw        &\qw      &\qw      &\qw&\qw\\
            \lstick{T}      &\targ   &\qw     &\qw      &\qw     &\qw       &\qw     &\targ   &\qw      &\qw        &\qw      &\qw      &\qw&\qw\gategroup{1}{1}{11}{5}{.7em}{--}\gategroup{3}{6}{7}{7}{.7em}{--}\\
            }\]
        \caption{Good Decomposition}
    \end{subfigure}
    \quad\quad\quad
    \begin{subfigure}[b]{0.4\textwidth}
    \[ \Qcircuit @R=1em @C=0.6em @!R=0.5em {
            \lstick{C_1}    &\qw     &\qw     &\qw      &\ctrl{2}&\qw       &\qw     &\qw     &\qw      &\qw        &\ctrl{2} &\qw      &\qw&\qw\\
            \lstick{C_2}    &\qw     &\qw     &\qw      &\ctrl{1}&\qw       &\qw     &\qw     &\qw      &\qw        &\ctrl{1} &\qw      &\qw&\qw\\
            \lstick{A_1^*}   &\qw     &\qw     &\ctrl{2} &\targ   &\ctrl{2}  &\qw     &\qw     &\qw      &\ctrl{2}   &\targ    &\ctrl{2} &\qw&\qw\\
            \lstick{C_3}    &\qw     &\qw     &\ctrl{1} &\qw     &\ctrl{1}  &\qw     &\qw     &\qw      &\ctrl{1}   &\qw      &\ctrl{1} &\qw&\qw\\
            \lstick{A_2^*}   &\qw     &\ctrl{3}&\targ    &\qw     &\targ     &\ctrl{3}&\qw     &\ctrl{3} &\targ      &\qw      &\targ    &\ctrl{3}&\qw\\
            \lstick{C_4}    &\qw     &\ctrl{2}&\qw      &\qw     &\qw       &\ctrl{2}&\qw     &\ctrl{2} &\qw        &\qw      &\qw      &\ctrl{2}&\qw\\
            \lstick{C_5}    &\qw     &\ctrl{1}&\qw      &\qw     &\qw       &\ctrl{1}&\qw     &\ctrl{1} &\qw        &\qw      &\qw      &\ctrl{1}&\qw\\
            \lstick{A_3^*}   &\ctrl{3}&\targ   &\qw      &\qw     &\qw       &\targ   &\ctrl{3}&\targ    &\qw        &\qw      &\qw      &\targ&\qw\\
            \lstick{C_6}    &\ctrl{2}&\qw     &\qw      &\qw     &\qw       &\qw     &\ctrl{2}&\qw      &\qw        &\qw      &\qw      &\qw&\qw\\
            \lstick{C_7}    &\ctrl{1}&\qw     &\qw      &\qw     &\qw       &\qw     &\ctrl{1}&\qw      &\qw        &\qw      &\qw      &\qw&\qw\\
            \lstick{T}      &\targ   &\qw     &\qw      &\qw     &\qw       &\qw     &\targ   &\qw      &\qw        &\qw      &\qw      &\qw&\qw\\
            }\]
        \caption{Bad Decomposition}
    \end{subfigure}
    
    \caption{Two possible decompositions of the Generalized Toffoli with $n=7$ controls and $m=4$ dirty ancilla. In (a), we get a Toffoli depth of 26 while (b) has a Toffoli depth of 30. This difference is owed to the fact that in (a) we have a single dirty $C^4X$ gate in the up section, while in (b) we use two $C^3X$ and one $C^3X$ in the up and down sections respectively. In general, it is more beneficial to maximize the number of Toffoli gates in each section. We have labeled the first up and down sections in (a).} 
    \label{fig:goodvsbad_dirty}
\end{figure}
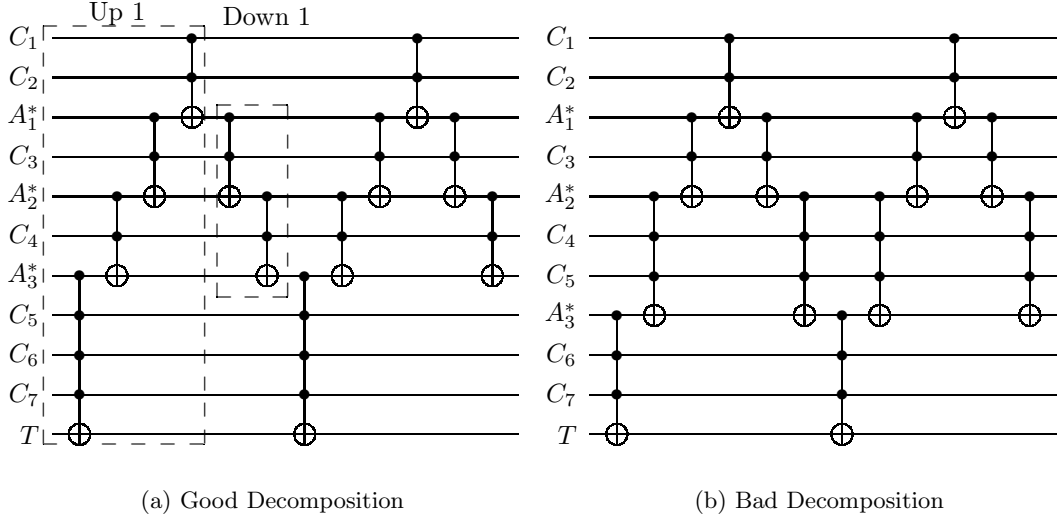

We define a Generalized Toffoli on a group as a Toffoli using all but the last element as controls and the last element as the target. As noted above, we need to perform up, down, up, down sections. An up section consists of performing Generalized Toffoli gates on the groups in order $G_{m+1}, ..., G_1$. A down section consists of performing Generalized Toffoli gates on the groups in order $G_2$, ..., $G_m$. This completes the decomposition into Generalized Toffoli gates. We can then proceed to decompose each of the Generalized Toffoli gates using the $k-2$ dirty ancilla decomposition, for which we've guaranteed there exist, resulting in a circuit composed only of Toffoli gates. Figure \ref{fig:goodvsbad_dirty} shows the results of two possible decompositions, with a decomposition maximizing the number of 2-control gates giving the best decomposition. Finally, we apply a Toffoli decomposition to get the final circuit.

\section{The Multi-target Gate}
The Generalized Toffoli is sometimes known as a ``multi-controlled-U'' gate where we have $n$ controls and we apply the $U$ gate if and only if each of the $n$ controls is a $1$. This gate depends on interacting all of the information together to decide if the $U$ gate should be applied or not. In contrast, the multi-target gate is one with a single control and $n$ targets, and we want to apply the $U$ gates to each of the targets if the control is on. This problem is much simpler than the previous one, but we present it here for completeness.

\subsection{Clean Ancilla}
Given $m$ clean ancilla, our control, and multiple targets. The idea is to copy the state of the control onto the clean ancilla and then apply the controlled single qubit operation $U$ using either the original control or one of the ancilla as the control. In this sense, the ancilla are treated as temporary controls.

Without any ancilla, we can decompose this gate trivially into $n$ single controlled $U$ operations. With too many ancilla, then copying the state of the control onto the ancilla could take more time than just a trivial decomposition. When we write to our $m$ ancilla using CNOT gates from the original control, we require depth $m$ operations. Then, we require depth $\left\lceil n/(m+1)\right\rceil$ to apply the controlled $U$ gates. Finally, we apply CNOT gates again in the same way to restore our ancilla. This is an upper bound on the depth because some of the $U$ gates can be done in parallel while the control is still writing to the ancilla. Still, trying to minimize this function, we find that it minimizes when $m=\sqrt{n/2}-1$. For this reason, we assert the number of ancilla used is less than or equal to $\sqrt{n/2}-1$. Also, we check if $2m + n/(m + 1) \le n$ and if not, we just use our trivial 0 ancilla decomposition.

\subsection{Dirty Ancilla}
Given $m$ dirty
ancilla, our scheme is generally the same as with clean. However, we must apply our $U$ gate twice when controlling by a dirty ancilla since its state is unknown. Here, we bound the depth of the circuit by $2m + 2n/(m+1)$. The optimal number of ancilla is given as $m = \sqrt{n} - 1$, the minimum of the equation above. There is a limit on the amount of benefit to be gained via dirty ancilla, and when there are too many, we simply fall back to a number which guarantees benefit i.e the number of used ancilla is given as $\min{\set{\sqrt{n} - 1, m}}$. 

\section{Analysis of Decompositions}

\begin{figure}
     \input{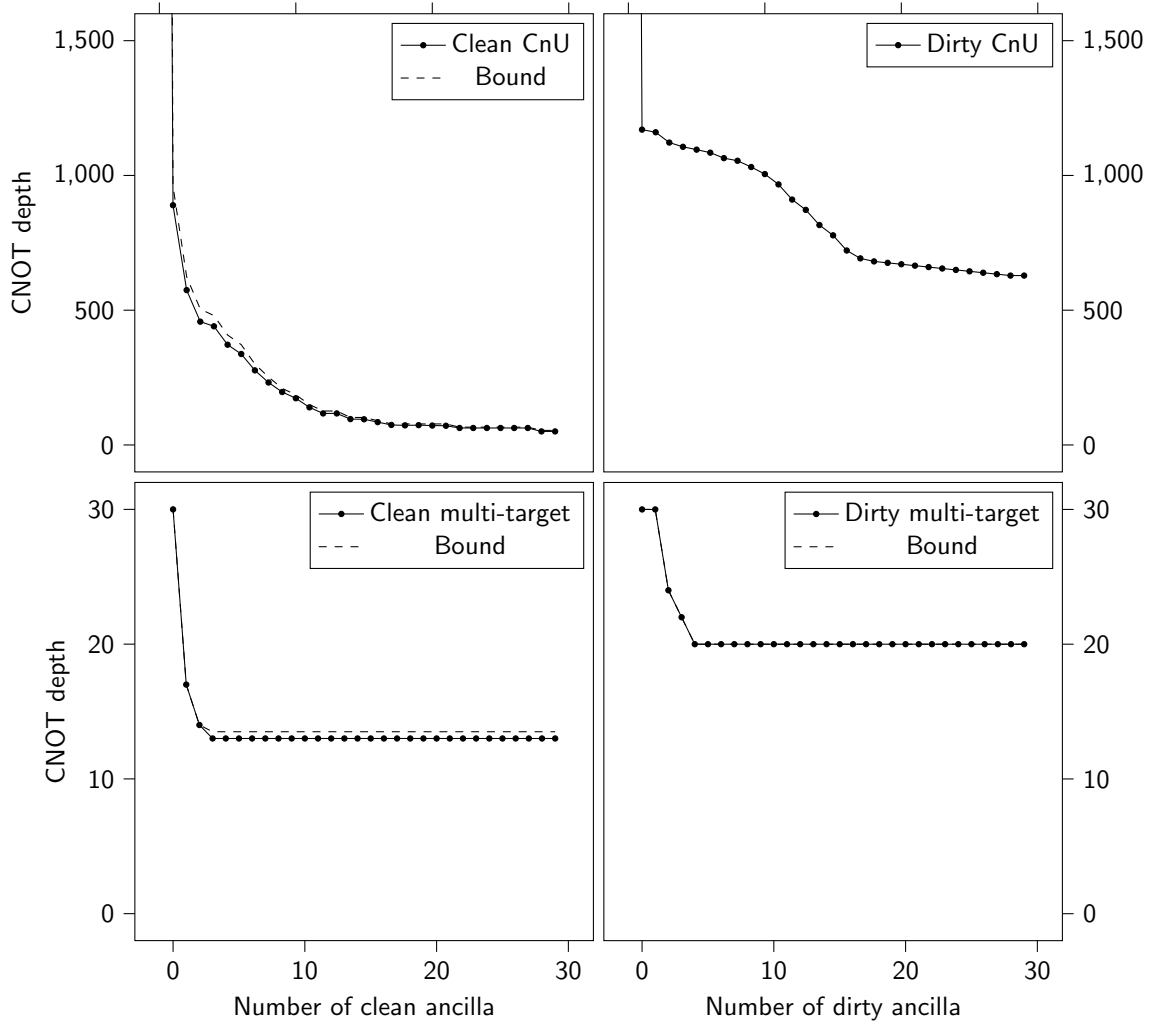}
     \caption{Depth of circuits for 30 controls/targets with varied number of ancilla. Our upper bound approximation well approximates the actual depth of the decompositions.  The zero ancilla decomposition of CnU is very bad with a CNOT depth of 8490 (not pictured). For the Dirty CnU gate, no upper bound is used to guide the decomposition. In this instance of Dirty multi-target, the upper bound is equivalent to the depth.}
     \label{fig:cost_plot}
\end{figure}

We reiterate that the cost metric used to inform the dynamic programming algorithm is an upper bound on the Toffoli depth and therefore on the actual depth. This is an upper bound because we consider the decomposition in layers, but when the circuits are actually pieced together, some gates near the boundaries may be able to be executed in parallel, reducing the overall depth slightly. We analyze how well our upper bound approximation is in Figure \ref{fig:cost_plot}. In general, our upper bound well approximates the resulting depth of the circuit. In the case of the Dirty $C^nU$ gate, we have no upper bound approximation. 

In each of the four decompositions provided, for the specified range of ancilla (clean or dirty), each guarantees either a reduction in depth or no change in the depth for each additional ancilla provided. Our primary cost metric was depth, rather than the gate count. Future work could include optimizing gate count as the primary cost metric. 

Clean ancilla provide more improvement than dirty ancilla where the additional space comes at a greater cost in gate count and depth. This cost does not always outweigh the potential benefit to circuit depth, as there are regimes where dirty ancilla provide substantial benefit. In both cases, however, the most improvement is obtained with the addition of the first few ancilla. The amount of marginal benefit quickly drops off, but, when available (up to the specified limits above) the ancilla should be used to minimize depth of a program.

The clean ancilla decomposition we present decreases in depth as more ancilla are given. However, for both Generalized Toffoli decompositions, there seems to be a middle region where we do not gain much marginal Toffoli depth reduction for each additional ancilla added. We speculate this could be due to the direct division we have between our two different schemes in our dynamic programming algorithm. We will always use scheme 1 until we are forced to do scheme 2. There might be a way to generalize the schemes to gain more from additional ancilla in the middle region; this is future work. 

We have presented decompositions of these gates for an arbitrary number of clean ancilla or an arbitrary number of dirty ancilla but not for some combination. It may be possible to combine these two approaches into a single unified algorithm to decompose a Generalized Toffoli or a multi-target gate for $m_1$ clean ancilla \textit{and} $m_2$ dirty ancilla.

\section{Conclusion}
We have presented 4 decompositions using an arbitrary number of clean or dirty ancilla. This bridges the gap between prior work which uses either only a large number of ancilla to achieve a logarithmic depth or few (0 or 1) ancilla. Our decompositions have room for improvement, but provide a generalized method for improving depth by the addition of clean or dirty ancilla.

Our variable ancilla decompositions may be useful for compiling circuits for quantum computer architectures with a limited number of qubits.  If multiple components of an algorithm run in parallel, the compiler may use this work to determine the optimal way to allocate ancilla to each Generalized Toffoli or multi-target gate used to minimize overall depth.

\section{Acknowledgements}
This work is funded in part by EPiQC, an NSF Expedition in Computing, under grants CCF-1730449/1832377, and in part by STAQ, under grant NSF Phy-1818914.n

\bibliography{ref}{}

\begin{thebibliography}{1}

\bibitem{Barenco}
Adriano Barenco, Charles~H. Bennett, Richard Cleve, David~P. DiVincenzo, Norman
  Margolus, Peter Shor, Tycho Sleator, John~A. Smolin, and Harald Weinfurter.
\newblock Elementary gates for quantum computation.
\newblock {\em Phys. Rev. A}, 52:3457--3467, Nov 1995.

\bibitem{GidneyBlogPost}
Craig Gidney.
\newblock Constructing large controlled nots, 2015.

\bibitem{He}
Y.~{He}, M.-X. {Luo}, E.~{Zhang}, H.-K. {Wang}, and X.-F. {Wang}.
\newblock {Decompositions of n-qubit Toffoli Gates with Linear Circuit
  Complexity}.
\newblock {\em International Journal of Theoretical Physics}, 56:2350--2361,
  July 2017.

\bibitem{ToffoliDecomp}
Vivek~V. Shende and Igor~L. Markov.
\newblock On the cnot-cost of toffoli gates.
\newblock 2008.

\end{thebibliography}
\bibliographystyle{plain}
\end{document}